% v1.5  --- released 25th August 1994 (M. Reed)
% v1.4  --- released 22nd February 1994
% v1.3  --- released  8th December 1992
%
% Copyright Cambridge University Press

\ifx\mnmacrosloaded\undefined \input mn\fi

\newif\ifAMStwofonts
%\AMStwofontstrue

\ifCUPmtplainloaded \else
  \NewTextAlphabet{textbfit} {cmbxti10} {}
  \NewTextAlphabet{textbfss} {cmssbx10} {}
  \NewMathAlphabet{mathbfit} {cmbxti10} {} % for math mode
  \NewMathAlphabet{mathbfss} {cmssbx10} {} %  "   "    "
  \ifAMStwofonts
    \NewSymbolFont{upmath} {eurm10}
    \NewSymbolFont{AMSa} {msam10}
    \NewMathSymbol{\upi}     {0}{upmath}{19}
    \NewMathSymbol{\umu}     {0}{upmath}{16}
    \NewMathSymbol{\upartial}{0}{upmath}{40}
    \NewMathSymbol{\leqslant}{3}{AMSa}{36}
    \NewMathSymbol{\geqslant}{3}{AMSa}{3E}

  \else
    \def\umu{\mu}
    \def\upi{\pi}
    \def\upartial{\partial}
  \fi
\fi

\input epsf

\def\postfig#1#2{                 % define a macro that puts in space if epsf
        \ifx\epsfbox\undefined    % is not available. #1=size, #2=file name
                \vskip 91mm
        \else
                \epsfxsize=#1\hfill\epsfbox{#2}\hfill
        \fi }

% Marginal adjustments using \pageoffset maybe required when printing
% proofs on a Laserprinter (this is usually not needed).
% Syntax: \pageoffset{ +/- hor. offset}{ +/- vert. offset}
% e.g.    \pageoffset{-3pc}{-4pc}

\pageoffset{-2.5pc}{0pc}

\loadboldmathnames

%\Referee   %  uncomment this for referee mode (double spaced)

% \pagerange, \pubyear and \volume are defined at the Journals office and
% not by an author.

%\onecolumn        % enable one column mode

% \letters          % for `letters' articles

\pagerange{xxx--xxx}    % `letters' articles should use \pagerange{Ln--Ln}
\pubyear{2001}
\volume{xxx}
% \microfiche{}     % for articles with microfiche
% \authorcomment{}  % author comment for footline

\begintopmatter  %  start the two spanning material

\title{Multifrequency studies of the enigmatic gamma-ray source 3EG J1835+5918}
\author{O.~Reimer$^{1,2}$, K.T.S.~Brazier$^{3}$, A.~Carrami\~nana$^{4}$,
G.~Kanbach$^{5}$, P.L.~Nolan$^{6}$, and D.J.~Thompson$^{1}$}
\affiliation{$^1$ NASA Goddard Space Flight Center, Code 661,
Greenbelt MD 20771, USA}
\affiliation{$^2$ NAS/NRC Research Associate}
\affiliation{$^3$ University of Durham, South Road, Durham DH1 3LE, England}
\affiliation{$^4$ Instituto Nacional de Astrofisica Optica y Electronica,
Tonantzintla, Mexico}
\affiliation{$^5$ Max-Planck-Institut f\"ur extraterrestrische Physik,
85740 Garching, Germany}
\affiliation{$^6$ W.W.~Hansen Experimental Physics Laboratory, 
Stanford University, Stanford CA 94305, USA}

\shortauthor{O. Reimer}
\shorttitle{Multifrequency studies of 3EG J1835+5918}

\acceptedline{Accepted 2001 February 8. Received 2001 January 19; 
in original form 2000 August 15}

\abstract {The EGRET telescope aboard NASAs Compton GRO has repeatedly
detected 3EG J1835+5918, a bright and steady source of high-energy gamma-ray 
emission which has not yet been identified. The absence of any likely 
counterpart for a bright gamma-ray source located 25$^\circ$ off the 
Galactic plane initiated several attempts of deep observations at other
wavelengths. We report on counterparts in X-rays on a basis of a 
60 ksec ROSAT HRI image. In order to conclude on the plausibility of the 
X-ray counterparts, we reanalysed data from EGRET at energies above 100 MeV 
and above 1 GeV, including data up to CGRO observation cycle 7. 
The gamma-ray source location represents the latest and probably the
final positional assessment based on EGRET data. We especially address the
question of flux and spectral variability, here discussed using the largest and 
most homogeneous data set available at high-energy gamma-rays for many years.
The results from X-ray and gamma-ray observations were used in a 
follow-up optical identification campaign at the 2.2 m Guillermo Haro Telescope at
Cananea, Mexico. VRI imaging has been performed at the positions of all of 
the X-ray counterpart candidates, and spectra were taken where applicable. 
The results of the multifrequency identification campaign toward this enigmatic
unidentified gamma-ray source are given, especially on the one object which 
might be associated with the gamma-ray source 3EG J1835+5918. This object 
has the characteristics of an isolated neutron star and possibly of a
radio-quiet pulsar.}


\maketitle

\section{Introduction}

The high-energy gamma-ray source 3EG J1835+5918 (also known as GRO J1837+59, 
2EG J1835+5919, and GEV J1835+5921) has been subject of considerable interest
since its discovery early in the CGRO mission by EGRET (Nolan et al. 1994).
Unidentified sources at high-energy gamma-ray wavelengths present significant 
challenges today. The large error boxes preclude any simple identification based 
on positional association only. The absence of gamma-ray lines eliminates the 
possibility of making direct redshift or source composition measurements. In most 
cases, the photon statistics are inadequate to carry out conclusive periodicity 
searches. Even long-term time variability is hard to establish unless strong 
flaring behavior is seen. Multiwavelength studies have been one of the most useful 
identification techniques. The blazar-class Active Galactic Nuclei (AGN) seen by 
EGRET, for example, are characterized by strong radio emission (typically 
$\sim 1$ Jy at 5 GHz) and a spectral energy distribution (SED) with one 
peak at IR-UV frequencies and a second peak in the gamma-ray band. Gamma-ray 
pulsars have high $F_\gamma$/$F_{radio}$ and often $F_\gamma$/$F_X$ 
ratios, with a single clear SED peak in the hard X-ray to
gamma-ray range. The energy requirements of gamma-ray production
demand powerful sources; therefore ordinary stars and most normal galaxies
can almost certainly be ruled out as gamma-ray sources. 

3EG J1835+5918 is a bright gamma-ray source discovered at high Galactic latitudes 
by EGRET, and this source was not until recently identified with any plausible
counterpart at other wavelengths. This has to be seen in respect to the observational
fact that, other than 3EG J1835+5918, all of the bright gamma-ray sources at
high Galactic latitudes have proved to be coincident with blazar-type AGN or 
flat spectrum radio quasars (FSRQ). However, even the most recent correlation of 
gamma-ray sources from the Third EGRET Catalogue (Hartman et al. 1999) with radio 
observations at 4.85 GHz from the Greenbank survey does not suggest any radio 
counterpart for 3EG J1835+5918 (Mattox et al. 2001). A dedicated deep search for 
pulsed radio emission with the Greenbank telescope at 770 MHz towards 
3EG 1835+5919 yielded only a 1 mJy flux density upper limit 
(Nice $\&$ Sawyer 1997). At the neighboring wavelengths in gamma-rays, 
observations have yielded only upper limits for an object at the position of 
3EG J1835+5918: the Whipple Telescope (E $> 300$ TeV) from observations made in 
1993 (Kerrick et al. 1995), and COMPTEL (0.75 - 30 MeV) throughout their total 
exposure from 7 years of CGRO observations (Sch\"onfelder et al. 2000).
At the 5th Compton Symposium (Reimer et al. 2000, Carrami\~nana et al. 2000),
we reported for the first time on X-ray counterparts for 3EG J1835+5918, suggesting
an indication of a possible radio-quiet pulsar. An independent search by Mirabal
et al. 2000 utilizing the same X-ray data, but separate optical observations,
concluded instead that 3EG J1835+5918 was likely to be a prototype of a new type
of high-energy gamma-ray source.\note {*}{Since this paper has been submitted, 
Mirabal and Halpern revised their earlier conclusion based on additional data, 
agreeing with the hypothesis that this source is most consistent with
being a radio-quiet neutron star.} 
Here we report our final analysis of the EGRET, X-ray, and optical data. Our
analysis reinforces our original conclusion that 3EG J1835+5918 has the 
characteristics of an isolated neutron star, possibly a Geminga-like
radio-quiet pulsar.

\begintable*{1}
\caption{{\bf Table 1.} $\gamma$-ray observations of 3EG J1835+5919.}
\halign{#\hfil & \quad \hfil#\hfil \quad & \hfil#\hfil \quad &
                       \hfil#\hfil \quad & \hfil#\hfil \quad & 
                                           \hfil#\hfil \quad & \hfil#\hfil\cr  
Viewing Period & Start Date & End Date & Aspect $
[^\circ]$ & $\sigma$ ($> 100$ MeV) & 
$\sigma$ ($> 1$ GeV) & comment \cr

0020 & 05/30/91 & 06/08/91 & 27.00 & 4.8 & 3.9 & outside 25$^\circ$ \cr
0092 & 09/12/91 & 09/19/91 & 28.61 & 4.4 & 4.5 & outside 25$^\circ$ \cr
0220 & 03/05/92 & 03/19/92 & 27.30 & 4.1 & 1.4 & outside 25$^\circ$ \cr
2010 & 11/17/92 & 11/24/92 & 23.28 & 5.1 & 1.2 & \cr
2020 & 11/24/92 & 12/01/92 & 21.46 & 6.3 & 5.2 & \cr
2030 & 12/01/92 & 12/22/92 & 26.55 & 9.3 & 5.1 & outside 25$^\circ$ \cr
2120 & 03/09/93 & 03/23/93 & 14.21 & 9.6 & 9.4 & \cr
3020 & 09/07/93 & 09/09/93 & 17.25 & 2.5 & u.l.& \cr
3032 & 09/22/93 & 10/01/93 & 17.25 & 7.5 & 6.7 & \cr
3034 & 10/01/93 & 10/04/93 & 22.07 & 3.3 & 4.8 & \cr
3037 & 10/17/93 & 10/19/93 & 17.25 & u.l.& u.l.& \cr
4030 & 11/01/94 & 11/09/94 & 28.78 & u.l.& u.l.& outside 25$^\circ$ \cr
7100 & 01/13/98 & 01/21/98 &  5.03 & 5.4 & 3.4 & \cr
7110 & 01/21/98 & 01/27/98 &  5.03 & 5.6 & 6.1 & \cr
co-added &        &          &       &  20.1 & 15.3 & \cr
}
\endtable

\section{The EGRET Source 3EG J1835+5918}

3EG J1835+5918 was first discovered at photon energies above 100 MeV by
the EGRET instrument aboard NASAs Compton GRO during regularly scheduled observations in 1991. 
The source was repeatedly seen whenever it was in the field of view of the EGRET instrument. 
However, the first observation performed with a close on-axis pointing towards 3EG J1835+5918 
using EGRET was only made in CGRO observation cycle 7 in 1998. EGRET observations performed at large 
off-axis viewing angle are problematic due to the degradation of the instrumental point spread function 
(PSF), most recently noted in Esposito et al. 1999. The first report on GRO J1837+59 
(Nolan et al. 1994) included only data from EGRET observations between 1991 and 1993. 
On the basis of six analyzed viewing periods a weak indication of flux variability was given. 
However, 3EG J1835+5918 was only once observed at less than 20$^\circ$ off-axis. 
The authors noted that, under those circumstances, a factor of two flux variation is not a 
strong indication for flux variability.
3EG J1835+5919 is listed in the Second EGRET catalogue of gamma-ray point 
sources (Thompson et al. 1995), which utilized a similar set of viewing
periods from CGRO observation cycles 1 and 2 as Nolan et al. 1994.
A separate variability study including EGRET data from cycles 1 to 3
again reported the source as being variable (McLaughlin et al. 1996), but the 
determined value of the variability criterion (V = 1.3) lies in the range where 
this parameter is rather inconclusive in deciding whether or not a source is 
indeed variable. With the appearance of the Third EGRET catalogue (E $> 100$ MeV) 
and the GeV source compilations (E $> 1$ GeV) (Lamb $\&$ Macomb 1997, Reimer et al.
1997), results from a total of 12 individual observations of 3EG J1835+5919 were 
reported. The source remains the brightest unidentified EGRET source 
outside the Galactic plane. Furthermore, it has been documented that there is 
gamma-ray emission from 3EG J1835+5919 up to the highest energies observable 
by EGRET.

In order to extend the coverage of 3EG J1835+5919 to its maximum we use for the 
first time all gamma-ray data taken by EGRET through the CGRO mission, including 
the previously unpublished observations made at a small off-axis angle in 
observation cycle 7. Table 1 lists the EGRET observations of 3EG J1835+5918. 

Throughout our analysis, we clearly distinguish between observations in which 
the angle between 3EG J1835+5918 and the instrument pointing direction was within 
or without 25$^\circ$. This distinction has been recommended by the EGRET 
instrument team for using the standard PSF (sources within 25$^\circ$ of the 
instrumental pointing) or using the wide-angle PSF if outside. 
The EGRET observations from CGRO observation cycle 7 extend significantly 
beyond the catalogued observations. They are separated by more than 3 years 
from the previous observations of 3EG J1835+5918. Both the
long-term observational aspect and the quality of the observation have
been improved: despite the lower efficiency of the EGRET spark
chamber, the 1998 observations were the first on-axis observations, unbiased 
from effects which most of the earlier EGRET observations of 3EG J1835+5918 
could suffer from. The narrow field-of-view mode, used in EGRET viewing
periods 710 and 711, does not introduce additional problems in this respect.

In order to determine the most likely location of the gamma-ray source,
we co-added the viewing periods using two different energetic thresholds 
(E $> 100$ MeV, and E $> 1$ GeV, respectively) and analyzed them individually.
Because 3EG J1835+5918 is a strong emitter up to GeV-energies, the best source
location was obtained from the highest energetic photons, 
where the instrumental PSF is significantly smaller than at lower energies. 
Using a likelihood method (Mattox et al. 1996), we determined the best position 
($> 1$ GeV) to be l = 88.80$^\circ$, b = 25.02$^\circ$, which is consistent with the 
position of 3EG J1835+5918 from our analysis above 100 MeV (l = 88.76$^\circ$, 
b = 25.09$^\circ$), the position given in 3EG and GEV catalogues, and the 
elliptical fit from Mattox et al. 2001 (l = 88.74$^\circ$, b = 25.08$^\circ$, 
a = 9.7$\arcmin$, b = 7.8$\arcmin$, $\Phi$ = 13$^\circ$). 
However, the additional data gave us positional errors of only 6$\arcmin$ and 8$\arcmin$ 
for the $68 \%$ and $95 \%$ confidence region, respectively. This enabled us to 
perform deep X-ray and optical studies of the entire region of 3EG J1835+5919. 
By using the position determined above 1 GeV, we examined the individual viewing
periods in order to evaluate the long-term characteristic of the gamma-ray
source flux. As discussed earlier, Nolan et al. (1994) and McLaughlin et al. 
(1996) indicated some source variability in 3EG J1835+5919 on the basis of smaller
data sets than presented here. However, the most recent variability study (Tompkins 1999) 
puts 3EG J1835+5918 clearly among the nonvariable sources, similar to the identified 
gamma-ray pulsars. Tompkins made use of an algorithm especially adopted for the 
characteristics of the observations by EGRET, i.e. sparse data sets from individual 
observations, often widely separated in time and characterized by different background 
levels. In addition, data from individual EGRET observations up to CGRO observation 
cycle were used in this study. A strict data selection (only within 25$^\circ$ on-axis)
among the gamma-ray observations was used, assuring a data set of comparable quality.
We complement the flux history of 3EG J1835+5918 with the data from 13-27 January 1998, 
the last high-energy gamma-ray data on this source to be taken for 
some years. 

Due to the generally lower efficiency of the EGRET spark chamber towards the end 
of the EGRET mission, the early 1998 viewing periods were evaluated using adjusted 
normalization factors (Esposito et al. 1999). These factors were checked 
quantitatively by means of a similar on-axis observation of Geminga during 
7-21 July 1998. Assuming that the instrumental sensitivity has not changed 
appreciably between these observations and that Geminga remains the stable 
gamma-ray emitter previously observed, this normalization for Geminga could 
be applied to the flux of 3EG J1835+5918 in the cycle 7 observations.
The light curve of 3EG J1835+5918 displaying all relevant EGRET observations is 
shown in Fig. 1, fluxes are for photon energies above 100 MeV, determined at the 
likelihood position of the GeV source. 
In addition, we consider (and label) observations with up to 25$^\circ$ off-axis 
separately from observations outside 25$^\circ$. The fluxes above 100 MeV and above 
1 GeV appear to be linearly correlated, considering the uncertainties in individual 
viewing periods arising from photon  statistics, especially for the sparse data of 
the detections above 1 GeV. 

We conclude, there is no indication of flux variability after all for 3EG J1835+1918, 
neither above 100 MeV nor above 1 GeV. Given the differing quality of the EGRET 
observations within their statistical and systematical uncertainties, we confirm 
the result from Tompkins (1999) and 
find 3EG J1835+5918 compatible with a nonvariable source of an average flux
of 5.9 $\times$ 10$^{-7}$ cm$^{-2}$ s$^{-1}$ (E $> 100$ MeV).

\beginfigure{1} \postfig {3truein}{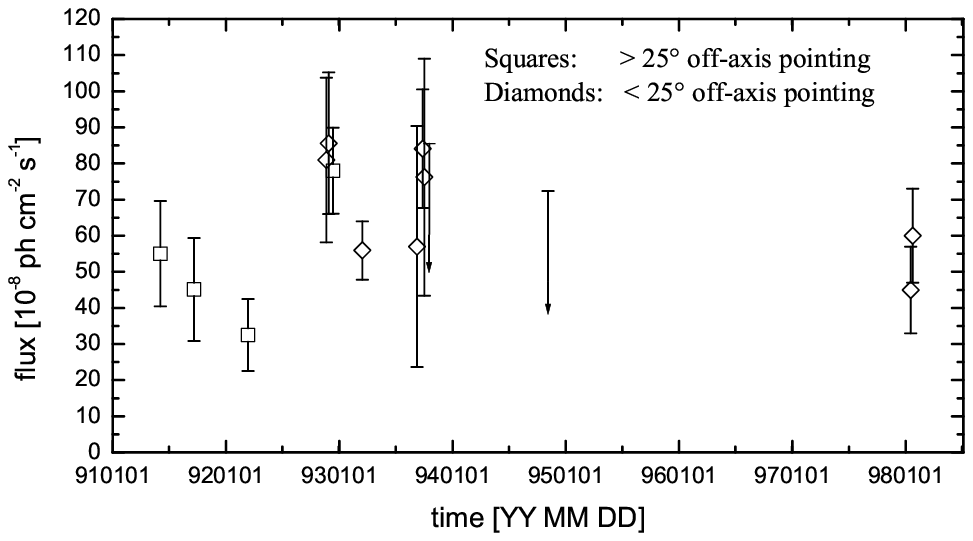}
%\vskip 30mm
\caption{{\bf Figure 1.} Gamma-ray flux history of 3EG J1835+5918
The fluxes (E $>$100 MeV) of 3EG J1835+5918, derived from data taken between 
1991 and 1998. To assure comparable quality in the data, observations 
within or without 25$^\circ$ of the target direction are marked differently. 
The lightcurve is consistent with a constant source considering statistical 
and instrumental restrictions from observations by the EGRET high-energy telescope.
} 
\endfigure

After concluding that 3EG J1835+5918 is consistent with having constant 
gamma-ray flux throughout the EGRET mission, we still have to examine the 
issue of its spectral variability. Nolan et al. 1996 reported evidence for 
spectral variability between individual EGRET viewing periods. Apparently, 
no correlation between spectral index and flux was found.
Hence, we re-examined the EGRET data on 3EG J1835+5918 for indication 
of spectral variability. Individual spectra in each of the relevant viewing 
periods were determined by simultaneously analyzing likelihood excesses of 
3 $\sigma$ detection significance and above. We derived a flux value or upper 
limit in each of ten energy intervals (30 MeV to 10 GeV) using a 
likelihood method. In cases when poor count statistics gave a spectrum dominated
by upper limits, the ten energy intervals were recombined into four
(30-100, 100-300, 300-1000, $>$1000 MeV), followed by the appropriate 
determination of the spectral slope. Also, when the source position
determined from likelihood analysis of an individual observation 
differed from the GeV-position, both positions were individually 
considered for consequences for the resulting spectrum. None of them introduces
relevant modifications in the resulting spectral slope. Therefore, the 
determined individual spectra could be compared at the best level currently
achievable for an unidentified high-energy gamma-ray source.

We find that the spectra of 3EG J1835+5918 determined from individual 
viewing periods are fully compatible within their statistical and systematic 
uncertainties throughout the entire EGRET mission, see Figure 2. A single power law spectral 
index of 1.73 $\pm$ 0.07 is consistent within 1 $\sigma$ for all individual spectra.

\beginfigure{2} \postfig {3truein}{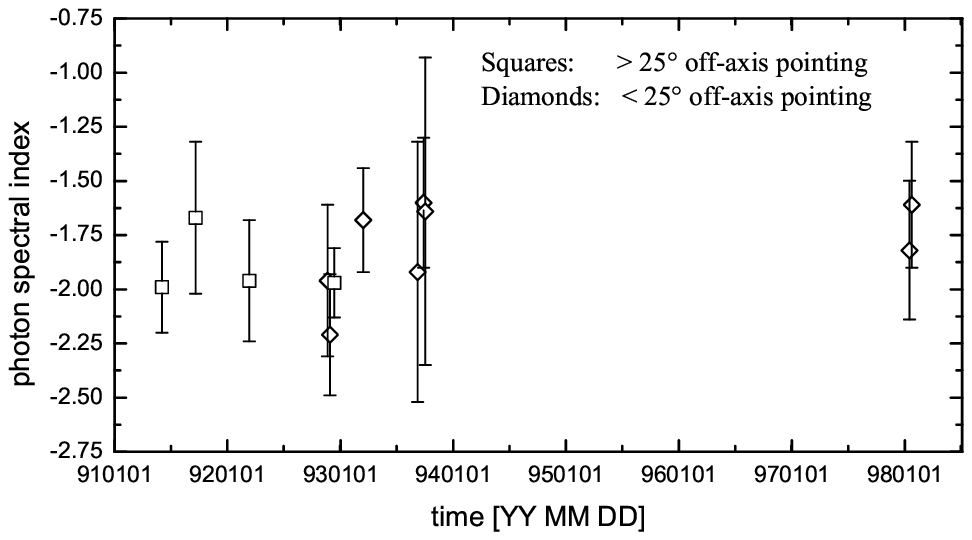}
%\vskip 30mm
\caption{{\bf Figure 2.} The gamma-ray spectral index of 3EG J1835+5918 between 1991 and 1998. 
The high-energy gamma-ray spectrum of 3EG J1835+5918 has been reported earlier to be variable. 
However, we find that the uncertainties in the determination of the spectrum only allow us to 
conclude, that the spectrum of 3EG J1835+5918 is still in agreement with being constant at a 
one sigma level throughout the entire EGRET coverage between 1991 and 1998. To emphasize 
comparable quality in the data, observations within or without 25$^\circ$ of the target 
direction are marked differently as in Figure 1.} 
\endfigure

With the consistency of the individual spectra throughout the EGRET
observations established, we co-added the data from cycles 1 to 7 in order to 
determine the best overall spectrum of 3EG J1835+5918. A single power-law fit appears to 
be inadequate for this source. The spectrum of 3EG J1835+5918 resembles the gamma-ray 
spectra of known gamma-ray pulsars like Vela or Geminga (Thompson et al. 1997)and the 
spectra of candidate gamma-ray pulsars like 3EG J0010+7309, as can be seen in Fig. 3: 
the hard power law spectral index, as determined to be -1.7 $\pm$ 0.06 between 
70 MeV and 4 GeV, the high-energy spectral cut-off or turnover as well as a possible 
spectral softening at the low energies. The restriction in energy when applying a single 
power law fit takes these features into account: it is based on the bins with the 
highest instrumental sensitivity.
The upper limits from COMPTEL (Sch\"onfelder 2000) do not constrain the shape of the spectrum
at lower energies. The TeV upper limits as reported by Kerrick et al. (1995) are consistent
with a rollover at 4 GeV, but certainly not with a simple extrapolation of the EGRET measured 
power law spectrum to the highest energies. 

\beginfigure{3} \postfig {3truein}{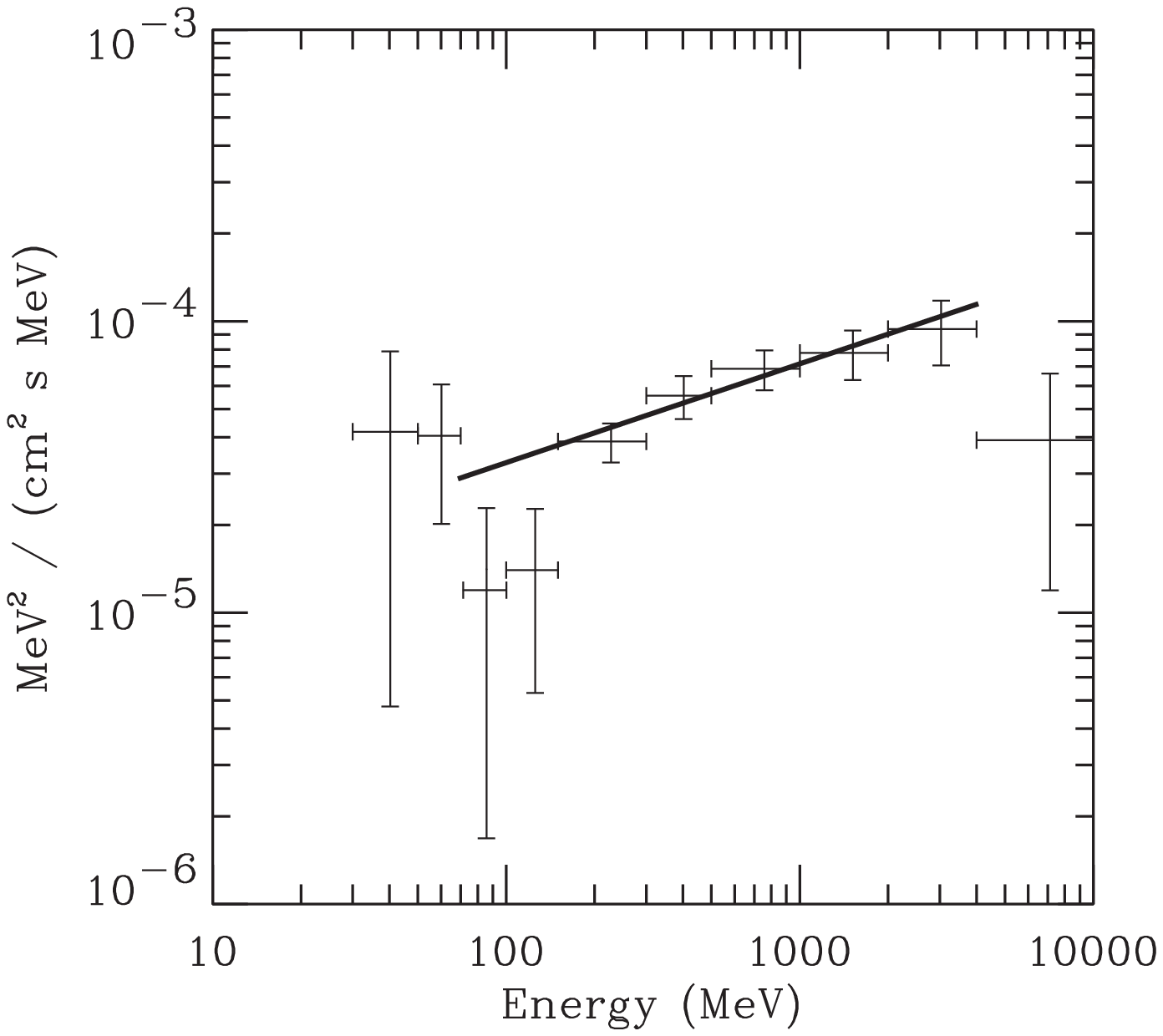}
%\vskip 30mm
\caption{{\bf Figure 3.} The high-energy gamma-ray spectrum of the source 
3EG J1835+5918, derived from data taken between 1991 and 1998. The power-law model fit was 
determined between photon energies of 70 MeV and 4 GeV, where this fit was 
applicable. Deviations from a single power-law model are seen at the lowest 
and highest energies.} 
\endfigure

\section{X-ray observations of 3EG J1835+5918}

The first observation towards 3EG J1835+5918 took place in February 1995.
A 9 ksec ROSAT HRI observation was performed, which reached a minimum 
detectibility limit of about 8 $\times$ 10$^{-14}$ erg cm$^{-2}$ s$^{-1}$ 
in the 0.1 to 2.4 keV band. A longer HRI observation was taken in 
December 1997/January 1998 as proposed by us in the ROSAT guest observer 
program. It exceeded the exposure of the former HRI observation by a factor of 
six with a total of 61.269 kseconds. Assuming a power-law spectrum with a photon 
index of -2 and a Galactic hydrogen column density of 
$N_H$=5 $\times$ 10$^{20}$ cm$^{-2}$, the limiting unabsorbed X-ray flux is 
about 2 $\times$ 10$^{-14}$ erg cm$^{-2}$ s$^{-1}$. Table 2 lists the detected 
X-ray sources in the long ROSAT HRI observation before astrometric corrections 
were applied. The astrometric correction is discussed in the context of reliable 
optical counterparts, see next section. Figure 4 shows the ROSAT HRI image, with 
the individual sources marked. Overlaid is the gamma-ray source location contour, 
determined solely from photons with energies above 1 GeV. The contours represent 
the 68$\%$ and 95$\%$ likelihood boundaries of the source location.

Only three of the ten sources found in this deeper HRI observations were detected in the 
earlier, short HRI observation (objects 1,3,9). We only note this here because the one object 
of further interest after investigating the deep HRI image, RX~J1836.2+5925, is rather close 
to the detection limit of the short observation. When examining X-ray variability for this 
particular source, it is generally hard to conclude based on two detections only, 
especially with one relatively close to its detection limit. 
However, any report on X-ray source variability would push the interpretations rather hard 
into some unique direction. Therefore we make no conclusions on the variability until further 
observations have been made and analyzed. 

The most recent X-ray observation took place between April 20-22, 1998, performed by 
ASCAs GIS and SIS detectors. The SIS data did not detect any object in the vicinity 
of 3EG J1835+5918, and the detected sources from the stacked GIS-images are located 
outside the GeV-source location we consider. 
There are no constraints from the nondetection of X-ray sources in the SIS images 
compared to the sensitivity achieved from the deep HRI observation. 

\beginfigure{4} \postfig {3truein}{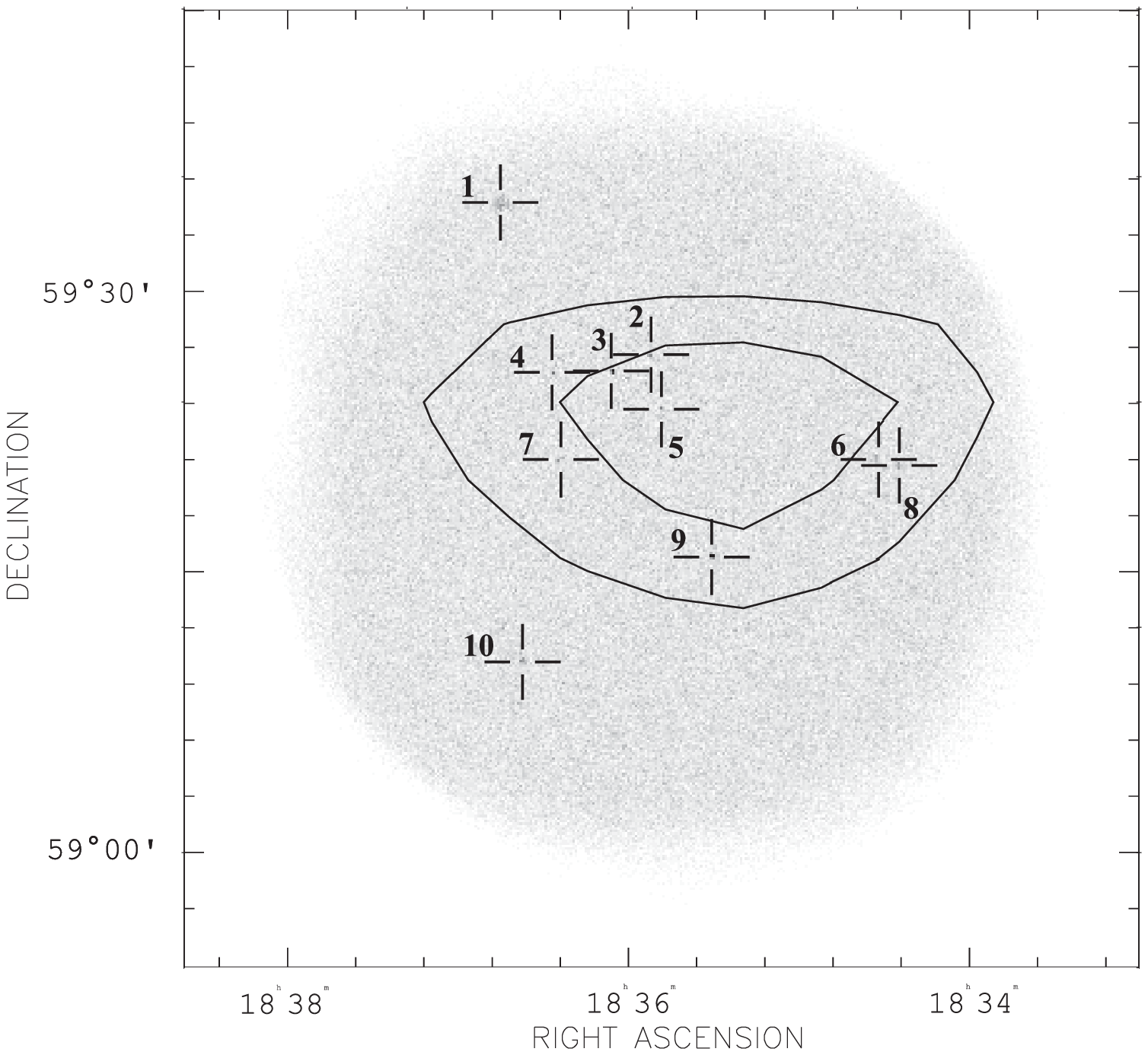}
%\vskip 30mm
\caption{{\bf Figure 4.} The long ROSAT HRI (0.1 - 2.4 keV) observation of the 
field of 3EG J1835+5918 from December 1997/January 1998. The X-ray image is overlaid 
with source location contours ($68\%$ and 95$\%$) of the high-energy gamma-ray source, 
determined above 1 GeV. The detected X-ray sources are indicated and were subject of a 
optical follow-on identification campaign.} 
\endfigure

\begintable*{2}
\caption{{\bf Table 2.} Sources detected in the 60 ksec ROSAT HRI X-ray observation and
and their relation to 3EG J1835+5918}
\halign{#\hfil & \quad \hfil#\hfil \quad & \hfil#\hfil \quad & 
                                         & \hfil#\hfil \quad & \hfil#\hfil \cr  
No & Name & RA & DEC & Identification & Association \cr
   &      & (J2000) & (J2000) & & with 3EG J1835+5918 \cr

 1 & RX~J1837.0+5934 & 18h 37m 00.82s & +59$^\circ$ 34$\arcmin$ 20.5$\arcsec$ & QSO (z $\simeq 0.47$) & highly unlikely \cr
 2 & RX~J1835.9+5926 & 18h 35m 58.09s & +59$^\circ$ 26$\arcmin$ 18.0$\arcsec$ & QSO (z = 1.87)        & unlikely \cr
 3 & RX~J1836.2+5925 & 18h 36m 13.62s & +59$^\circ$ 25$\arcmin$ 28.9$\arcsec$ & n.a.                  & candidate \cr
 4 & RX~J1836.6+5924 & 18h 36m 38.45s & +59$^\circ$ 25$\arcmin$ 24.0$\arcsec$ & QSO (z = 1.75)        & unlikely \cr
 5 & RX~J1835.9+5923 & 18h 35m 53.32s & +59$^\circ$ 23$\arcmin$ 29.0$\arcsec$ & QSO (z = 1.86)        & unlikely \cr
 6 & RX~J1834.4+5920 & 18h 34m 23.82s & +59$^\circ$ 20$\arcmin$ 51.0$\arcsec$ & M5V star              & highly unlikly \cr
 7 & RX~J1836.6+5920 & 18h 36m 36.78s & +59$^\circ$ 20$\arcmin$ 40.6$\arcsec$ & QSO (z = 1.36)        & unlikely \cr
 8 & RX~J1834.2+5920 & 18h 34m 14.23s & +59$^\circ$ 20$\arcmin$ 24.5$\arcsec$ & G7V star              & highly unlikly \cr
 9 & RX~J1835.5+5915 & 18h 35m 32.33s & +59$^\circ$ 15$\arcmin$ 39.3$\arcsec$ & M star                & highly unlikely \cr
10 & RX~J1836.8+5910 & 18h 36m 50.81s & +59$^\circ$ 10$\arcmin$ 03.7$\arcsec$ & K5V star              & highly unlikely \cr
}
\endtable

\section{Optical identifications in the vicinity of 3EG J1835+5918}

We studied the X-ray sources found in the vicinity of 3EG J1835+5918 for 
optical counterparts. If objects at optical wavelengths appear coincident 
at positions of the X-ray sources, the astrometry of the obtained ROSAT 
HRI-image could be verified and corrected. Deep follow-up observations have been 
conducted in order to conclude on the nature of those optical counterparts. 
Starting with a general assessment on the basis of DSS-2 plates and USNO-A2.0
catalogue listings, we subsequently performed VRI imaging of the entire 
field of the gamma-ray source at the Observatorio Astrof\'{\i}sico Guillermo 
Haro, located in Cananea, Sonora (lat=31$^\circ$, long=110$^\circ$ W). 
We used the Faint Object Spectrograph Camera (LFOSC) of the Landessternwarte 
Heidelberg, specially designed for optical counterpart identifications. The 
instrument allows photometric BVRI imaging in a 10$\arcmin \times 6\arcmin$ FOV and 
two modes of low-resolution spectroscopy. 
Observations in July and September 1997, and May 1998 were devoted to VRI 
imaging. When the ROSAT HRI image taken at January 1998 had been delivered and
thoroughly analyzed, spectroscopy was made in June and October 1999 by using the 
low resolution mode at 4200{\AA} to 9000{\AA} with a $\sim$8.3\AA/pixel 
sampling.

Fortunately, the search for counterparts has revealed one excellent astrometrical
measure in the field, a star listed in the Tycho-2 (3917 00934 1) and ACT-catalogues. 
The offset of the coincident X-ray object 8 (RX J1834.2+5920) is about 0.5 s in right ascension 
and 1.1$\arcsec$ in declination. A similar offset has been found at the USNO-A2.0 listed 
object coincident with the X-ray object 5 (RX J1835.9+5923). Offsets at other object pairs 
differ more, significantly at the edge of the field of view. Therefore we do not average any 
additional, in some cases contradictory offset parameter. We apply the correction found 
appropriate for object 8 and 5 throughout the entire HRI image. The following objects were
studied:

\beginlist
\item (1) {\bf RX~J1837.0+5934}: This X-ray source is outside the 68\% and 95\% GeV error contour, 
already signalling that an association is unlikely. Two USNO-A2.0 listed objects are positionally 
consistent with the X-ray source position.
The optical spectrum of the brighter object shows a strong emission line at 7198{\AA} which 
can be identified as redshifted H$\beta$, allowing us to associate the emission lines at 
6364{\AA} and 6012{\AA} with reshifted H$\gamma$ and H$\delta$. This quasar at $z\simeq 0.466$ is 
a very plausible counterpart for the X-ray source.

\item (2) {\bf RX~J1835.9+5926:} An optical object, located 7 $\arcsec$ NE from the X-ray source 
position has been found at the Cananea images (V$>20.2$; R=20.5$\pm$0.7). Its final identification 
with a QSO at $z=1.87$ by Mirabal et al. 2000 is plausible.

\item (3) {\bf RX~J1836.2+5925:} Two faint objects are at 11 $\arcsec$ NW and 14 $\arcsec$ SE 
of the X-ray source location, not listed in the USNO-A2.0. Given the low uncertainty on the X-ray 
position, an association between either of them and the ROSAT source is doubtful. Our images
have a detection limit around $R\ga 21$, and no other object is found anywhere closer to the X-ray 
source position.

\item (4) {\bf RX~J1836.6+5924:} An optical object (V=19.0$\pm$0.2, R=19.5$\pm$0.2) lies
centered on the X-ray position. It has been identified with a QSO at $z=1.75$ by Mirabal et al. 2000.

\item (5) {\bf RX~J1835.9+5923:} The optical object nearly centered at the X-ray position  
(V=19.3$\pm$0.3, R=19.1$\pm 0.2$) shows a strong emission line, which we identify with CIV1550, 
recovering CIV1909 redshifted to 5469{\AA} (emission doublet) and Mg2798 redshifted to $\sim$8020{\AA}. 
This quasar with $z=1.865$ is very likely to be the X-ray emitting source.

\item (6) {\bf RX~J1834.4+5920:} An optical object is more than 10$\arcsec$ distant from the 
X-ray position for which we measured V=18.04$\pm$0.10 and R=18.04$\pm$0.10. Its spectrum indicates a 
late type star, probably M5V. This X-ray source is 1.3 $\arcmin$ from the bright star coincident with 
X-ray source number 8, whose glow complicates the detection of objects fainter than magnitude 20. 
The potential identification of the M5V star with the X-ray source would rule out their association 
with the 3EG J1835+5918. If one does not follow this identification scheme due to the source
location offset, we find no other optical counterpart for this object.
However, this would leave the question unanswered, why we observe no X-ray emission from this 
M5-star down to the detection limit of the ROSAT HRI image.

\item (7) {\bf RX~J1836.6+5920:} There is no optical object brighter than V$\sim$21 inside the error box. 
Mirabal et al. 2000 found an optical object at B=21.3 at z=1.36 by a UV-excess selection technique using 
the Hobby-Eberly-Telescope. Although, we cannot adopt their argument of rejection by positional 
inconsistency of RX J1836.6+5920 with the EGRET-source location contour, the optical and X-ray properties 
of this QSO would make it highly unlikely to be the counterpart of 3EG J1835+5918.

\item (8) {\bf RX~J1834.2+5920:} The star Tycho 3917009341 with V$\sim$ 9.4 is 
centered at the X-ray source position. The spectrum obtained indicates a late G dwarf star, 
probably G7V.

\item (9) {\bf RX~J1835.5+5915:} A $V=15.6$ magnitude object is about 5$\arcsec$ from 
the X-ray position, and therefore marginally consistent with the X-ray source. Its 
spectrum indicates a late M-type star.

\item (10) {\bf RX~J1836.8+5910} is outside the 68\% and 95\% location contour of 
the EGRET source, and their physical association is practically rejected on positional 
inconsistency. A bright star V$\sim$11th magnitude (Tycho 3917018071) is coincident with 
the X-ray position. Our spectrum indicates a K star, probably K5V.
\endlist

We reject only sources 1 and 10 primarily on the grounds of inconsistency with the GeV source location. 
If this argument does not absolutely disqualify both X-ray sources as candidates for an association with 
the gamma-ray source, their identifications probably do. Also, the identified coronal emitting stars do
not qualify as likely counterparts for 3EG J1835+5918. 
The four QSO-identifications (z=1.36, 1.75, 1.87, 1.86) certainly need a closer look.
First of all, these are not blazar-class AGN which constitute the vast majority of the known
EGRET QSO sources. Also, their redshifts would lie at the tail of the redshift-distribution of 
gamma-ray loud Active Galactic Nuclei (Mukherjee 1997). The lack of observed radio emission is 
an additional hint that they do not belong to the class of AGN that EGRET could actually see 
(Mattox et al. 2001). The obvious mismatch between one of the brightest gamma-ray sources at 
high Galactic latitudes and the lack of observable radio emission would exhibit 3EG J1835+5918 
as a unique source among the gamma-ray sources identified with active galactic nuclei.  
Secondly, the intrinsic gamma-ray emission characteristics argue against a quasar counterpart.
The EGRET-detected AGN are highly variable at all timescales currently observable. The average spectral 
index of the EGRET-detected AGN is significant softer than the one determined for 3EG J1835+5918,
in fact it would put 3EG J1835+5918 as the hardest source among the more than ninety
EGRET-detected AGNs. The lack of obvious source variability as well as the hard spectral index
of 1.7 $\pm$ 0.06 would solely argue against a quasar identification; both arguments together do so 
with a rather high degree of confidence. 
Parameters like the optical to X-ray and/or optical to $\gamma$-ray luminosity ratio for the 
QSO-counterparts do not give a unique signature. The range which could be occupied by QSOs in both 
parameter spaces is rather wide. Also, the correlation signatures reported for 61 of the gamma-ray 
loud blazars (Cheng et al. 2000) do not allow any additional conclusive information for the four 
X-ray sources identified as radio-quiet QSOs. Summarizing, the various observational facts concerning 
an association between each of the X-ray sources identified as quasars and the unidentified 
gamma-ray source 3EG J1835+5918 do not support such an interpretation. More exotic scenarios must 
be called in order to accept one of the quasar identifications, like hypothesized radio-quiet 
blazars (Mannheim 1993) or a $\gamma$-ray AGN with a shifted SED (Ghisellini 1999), where the 
synchrotron component will fall in the MeVs and the IC-component peaks at TeV-energies. 
Currently, observational constraints from COMPTEL and Whipple do not indicate that the SED peaks at 
any other wavelength band except where 3EG J1835+5918 is observable, bright and steady: the 30 MeV 
to 10 GeV band where EGRET operated for nine years. 

Therefore, we have to consider further only one X-ray source in the vicinity of 3EG J1835+5918,
RX J1836.2+5925 (object 3, see Table 2). The lack of an optical counterpart and therefore of the 
possibility to identify it by means of an optical spectrum keeps this source as the only
unidentified X-ray source which could have an association with the $\gamma$-ray source
3EG J1835+5918. Using their independent optical observation, Mirabal et al. 2000 also
concluded that RX J1836.2+5925 is the most probable X-ray counterpart to the gamma-ray source.

\beginfigure{5} \postfig {3truein}{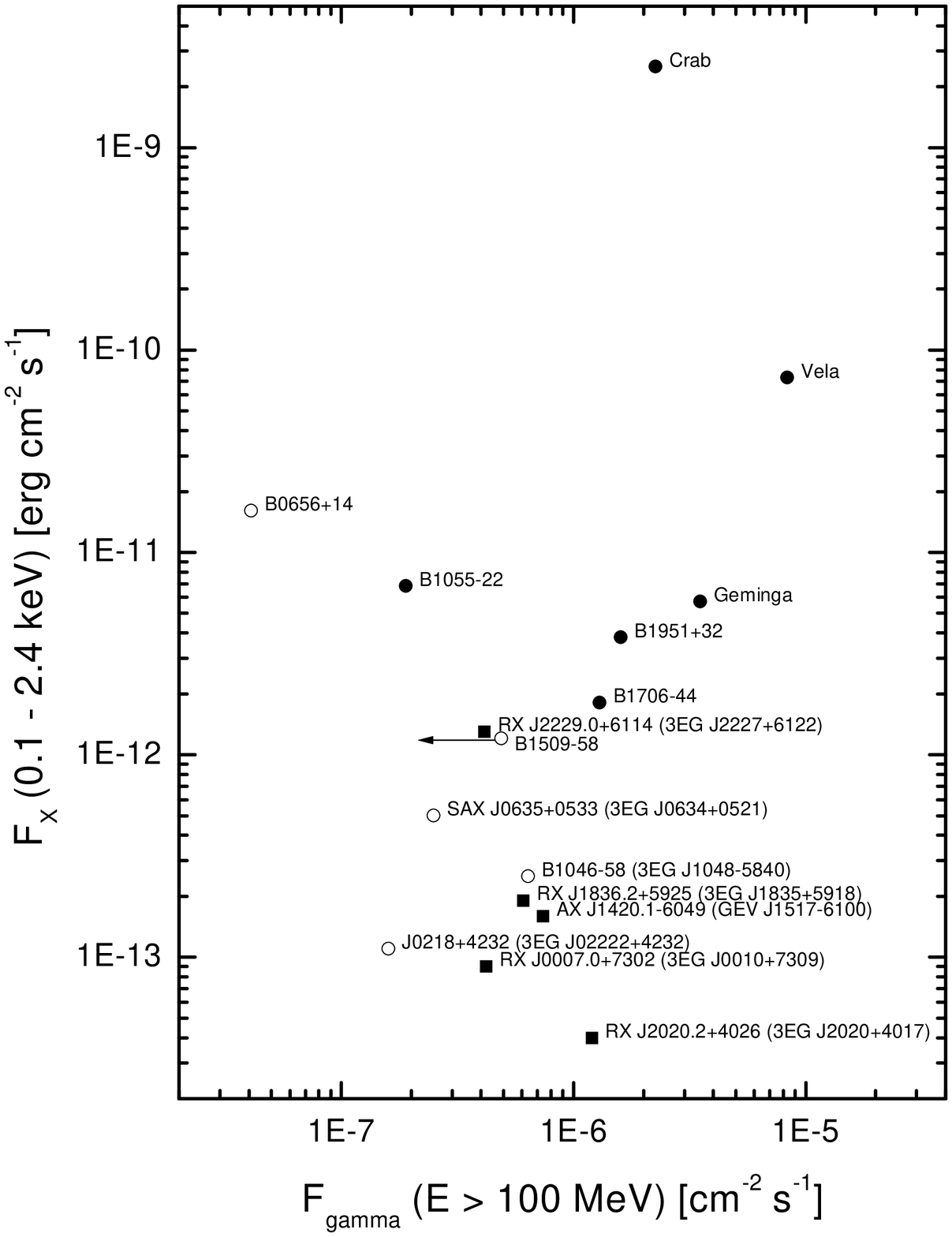}
%\vskip 30mm
\caption{{\bf Figure 5.} X-ray and gamma-ray fluxes of high-confidence 
pulsar detections (filled circles), probable associations between pulsars and 
high-energy gamma-ray sources (open circles), and candidate radio-quiet pulsars
(filled squares).
All X-ray fluxes are given for the 0.1 to 2.4 keV energy band, in cases of
different energy band quoted in the literature (Crab, B1509-58, B1951+32,
B1046-58 from Becker \& Tr\"umper 1997, SAX J0635+0533 from Kaaret et al. 1999, 
AX J1420.1-6049 from Roberts et al. 2000, J0218+4332 from Kuiper et al. 1998), 
the flux is normalized into the chosen energy band. The gamma-ray fluxes are given 
above 100 MeV, in cases where different event selection criteria were used 
(B0656+14: $> 50$ MeV (Ramanamurthy et al. 1995), B1046-58: $> 400$ MeV (Kaspi et al. 2000), 
B1509-58: 30 to 100 MeV (Kuiper et al. 1999), J0218+4332: 100 to 300 MeV (Kuiper et al. 2000))
the appropriate gamma-ray flux above 100 MeV has been determined for the 
energy band desired here. 
} 
\endfigure

\section*{Discussion}

3EG J1835+5918 is a persistent high-energy gamma-ray source located at  
high Galactic latitudes and has been observed repeatedly by EGRET. It is 
characterized by a hard power law and a spectral break or turn-over above 4 GeV. 
It appears to be a nonvariable source in terms of its flux as well as 
its spectral shape throughout the entire EGRET mission, despite suggestions of 
variability from earlier analyses. Its gamma-ray properties are typical of those
observed from gamma-ray pulsars like Vela or Geminga, and candidate radio-quiet 
pulsars like 3EG J2020+4017, 3EG J0010+7309, and 3EG J2227+6122.
Our deep ROSAT HRI observation revealed several X-ray sources consistent with 
the location of the observed GeV-emission of 3EG J1835+5918. As a result of the 
identification campaigns independently carried out by Mirabal et al. 2000 and ourselves, 
only one of the ten X-ray sources still attracts interest to be considered further for 
an association with the $\gamma$-ray source. This source, RX J1836.2+5925,
is characterized by an obvious lack of radio-emission, indetectibility by means
of an UV-excess identification technique, lack of optical counterpart up to
V$\sim$23 mag in the V- and B-bands, and location well inside the 68\% likelihood test 
statistic contours of 3EG J1835+5918. Our HRI observation contain no information on the 
X-ray spectrum of RX J1836.2+5925. Hence, assuming that this X-ray source is the most 
likely counterpart to 3EG J1835+5918, we are restricted to using the X-ray flux of 
RX J1836.2+5925 and the gamma-ray properties of 3EG J1835+5918
to investigate the characteristics of the object.

To do so, we can use its multifrequency properties to ascertain its characteristics. 
The high $F_\gamma$/$F_{radio}$ value seems to rule out a blazar origin. The already 
noted similarities in the gamma-ray characteristics with known gamma-ray pulsars 
(Thompson et al. 1997) or radio-quiet pulsar candidates (for a recent summary, see 
Brazier \& Johnston 1999) definitely suggest such a radio-quiet pulsar candidate. 
We have re-examined the gamma-ray and X-ray fluxes for all the known and candidate 
pulsars, using a consistent energy range in both bands.  
Comparing the flux of 3EG J1835+5918 in $\gamma$-rays (E $> 100$ MeV) and 
RX J1836.2+5925 in X-rays (0.12 - 2.4 keV), this source falls among the 
candidates currently considered for associations between gamma-ray sources 
and X-ray sources with proven or suspected neutron star origin, see Fig. 5. 
Nearly all of the candidate gamma-ray pulsars lie at the bottom end of the 
sensitivity feasible for the last generation of X-ray instruments like ROSAT, 
ASCA, and SAX. Obviously, only deep observations could reveal counterparts at 
all or with features not easily explained by any other astronomical objects. 
The lack of optical counterparts up to V$\sim$23 mag and radio emission for 
RX J1836.2+5925 is a further characteristic signature for isolated, radio-quiet 
neutron stars (Caraveo et al. 1996), and ideally demonstrated by Geminga as 
prototype (Bignami \& Caraveo 1996).
 
Although many of the candidate radio-quiet pulsars beside Geminga itself are 
located within or near SNRs, 3EG J1835+5918 does not. Neither radio observations
nor the X-ray data yield any hint of a SNR in the vicinity of this object, and
the high Galactic latitude seems to rule out the possibility of obscuration that 
might hide one.

If 3EG J1835+5918/RX J1836.2+5925 is not of quasar origin and also not the first 
candidate of an hypothesized extragalactic astronomical object bright and steady in 
gamma-rays, faint in X-rays, and yet undetectable at optical and radio wavelengths, 
it will reside within our Galaxy. We therefore have to suspect an isolated radio-quiet 
neutron star candidate. With Geminga as the only established pulsar of a predicted class 
of radio-quiet pulsars, extremely well characterized with its highly resolved high-energy 
lightcurve, independent measurements of rather weak radio emission and a faint optical counterpart
with noticeable proper motion, a comparison of observational parameters in analogy
with 3EG J1835+5918/RX J1836.2+5925 might be appropriate. First, Geminga is three times brighter
in gamma-rays and about fifty times brighter in X-rays. To extrapolate from the distance to 
Geminga (d $\ga 160$ pc, Caraveo et al. 1996) using the observed fluxes, 3EG J1835+5918 would 
lie between 250 pc (scaling from gamma-rays) and 1.1 kpc (scaling from X-rays), assuming
the same beaming as Gemingas in both cases. 
Besides, pulsars tend to begin their life in the Galactic plane. A pulsar moving with 
a typical velocity of about 350 km s$^{-1}$ would move only 300 pc
even in a lifetime of $10^6$ years, while an object seen at b = 25$^\circ$ would have to
move more than 420 pc from the plane if it were at a distance greater than 1 kpc. 
As pointed out by Yadigaroglu and Romani (1995) discussing
the beaming evolution of pulsars in an outer-gap model, the beaming fraction 
becomes rather small as the pulsars age increases. Therefore a distant but 
old pulsar would have to be immensely powerful or exceptionally beamed.
If 3EG J1835+5918 were more distant, then its gamma-ray luminosity would exceed that of
Geminga, but if it were closer, then the surface brightness in X-rays of the neutron star
would have to be lower than Gemingas. This indicates, that either the efficiency of the 
emission mechanism is different and/or the parameter space which radio-quiet pulsar 
candidates could occupy is wide spread. 
In contrast to energetic pulsars like Vela or B1706-44, nonthermal emission or pulsar nebular 
features have not been observed in the case of RX J1836.2+5925 so far. Nor is it an extended
source in the X-rays. The lack of an associated SNR as well as the rare chance to find a 
similar pulsar at such high Galactic latitude (i.e. nearby) argues against a young pulsar 
in the case of 3EG J1835+5918. 
However, the striking similarities in the gamma-ray properties between Geminga, other
candidate radio-quiet pulsars and 3EG J1835+5918, the absence of a radio and optical counterpart of 
RX J1836.2+5925, and, finally, the arrangement of 3EG J1835+5918/RX J1836.2+5925 
among the other candidate gamma-ray pulsars (Fig. 5) still leaves room for accepting a 
hypothesis being an older but radio-quiet pulsar. 

Halpern at al. 2000 hypothesized in the case of 3EG J2227+6122/RX J2229.0+6114 a medium aged 
gamma-ray pulsar population, efficient enough to be comparable to older pulsars like Geminga 
or B1055-52, although matching the luminosity (i.e. spin-down power) constraints from gamma-ray 
observations. With 3EG J2227+6122/RX J2229.0+6114 rather close in its $F_\gamma$/$F_X$ to young 
pulsars with nonthermal emission and pulsar nebular features, 3EG J1835+5918/RX J1836.2+5925 lies 
well among the other candidate radio-quiet pulsars. 
We rather stress the similarities seen in the $F_\gamma$/$F_X$ to existing candidate gamma-ray pulsars, 
together with its characterization by apparently similar gamma-ray properties as identified pulsars, 
however faint in X-rays and no counterpart yet at optical and radio frequencies. Mirabal and
Halpern (2001) have recently reached a similar conclusion.

With 3EG J1835+5918/RX J1836.2+5925 being in the range of a isolated neutron star candidate explanation, 
other suggestions still need to be looked at. A similarity to the widely discussed association between 
LSI +61$^\circ$303/3EG J0241+6103 and SAX J0635+533/3EG J0634+0521 does not seem to be appropriate for 
various and strict reasons. Both are binary systems (Be/X-ray), characterized by different states of 
variability on all but the shortest timescales. Also, they show a gamma-ray spectrum significantly 
softer than determined in the case 3EG J1835+5918. Most severe, since only a small number of such 
Be/X-ray binaries is expected (Vanbeveren et al. 1998), the high Galactic location of 3EG J1835+5918
would indicate a rather nearby one, definitely conflicting with the lack of any optical counterpart 
up to 23rd magnitude. With extremely high degree of confidence we can therefore rule out the 
probability of seeing a massive star/compact object binary system here.  

There have been other suggestions made for 3EG J1835+5918. Plaga et al. 1999 explicitly 
refer to 3EG J1835+5918 as an example for a hot-spot of a Galactic gamma-ray burst.
The low X-ray luminosity of RX J1836.2+5925 and the upper limit from the Whipple observation 
are severe observational arguments against such a hypothesis.
Isolated accreting black holes (Armitage \& Natarajan 1999) were suggested to be
energetically consistent with unidentified sources at all Galactic latitudes.
However, if their high energy emission occurs via similar processes to those in AGNs,
it is expected to observe variability on all timescales. This is definitely not the
case for 3EG J1835+1918. The persistent nature of the detections from 3EG J1835+5918 
also excludes any similarities to gamma-ray transient like GRO J1838-04.

Certainly, neither the X-ray data nor the gamma-ray data currently 
allow wide range period scans for pulsations without known ephemerides
(Jones 1998). A search for periodicity will have to be postponed
until more sensitive instruments like XMM in the X-rays or GLAST in
the gamma-rays have observed 3EG J1835+5918. However, if a restrictive
set of parameters can be predicted from pulsar models or if a lightcurve
can be derived from another wavelength, the archival EGRET data will
permit the discovery of pulsations in the gamma-rays.  The long
observational history presented here will certainly assist in any 
such effort. Finally, RX J1836.2+5925 might be identified as a 
neutron star by extremely deep optical imaging/spectroscopy. To unambiguously 
relate 3EG J1835+5918 to a known class of astronomical objects would be 
of extreme importance for any collective studies of gamma-ray sources,
as well as for studies of contributors to the diffuse gamma-ray
background, not to mention the gain for pulsar physics if the existence
of another isolated neutron star in gamma-rays is confirmed.

\section*{Acknowledgments}

A part of this work was performed while O.R. held a National Research Council/NASA GSFC 
Resident Research Fellowship. KTSB acknowledges the use of the Starlink facilities at the 
University of Durham, England. A part of this work was funded by CONACyT grant 25559-E (A.C.).
We acknowledge LSW Heidelberg for the use of the LFOSC instrument an Cananea. The Digitized 
Sky Surveys were produced at the Space Telescope Science Institute under U.S. Government 
grant NAG W-2166. This work made use of the USNO-A2.0 catalogue.

\appendix
\bye